\documentclass[preprint,5p]{elsarticle}

\usepackage{graphicx}
\usepackage{enumitem}
\usepackage{bbm}
\usepackage{relsize}
\usepackage{xcolor}
\usepackage[caption=false]{subfig}
\usepackage[pdftex]{hyperref}
\usepackage{amsmath,amssymb}
\newcommand{\Tr}{{\rm Tr}\,}

\journal{Computer Physics Communications}

\begin{document}
	
	\begin{frontmatter}
		
		\title{Explicit volume-preserving numerical schemes for relativistic trajectories and spin dynamics}
		
		\author[1]{Renan Cabrera}
		\ead{rcabrera@princeton.edu}
			
		\author[2]{Andre G. Campos}
		\ead{agontijo@mpi-hd.mpg.de}
	
		\author[3]{Denys I. Bondar}
		
		\author[4,5,6]{Steve MacLean}
		\ead{steve.maclean@emt.inrs.ca}	
		
		\author[4,5]{Fran\c{c}ois Fillion-Gourdeau\corref{cor1}}
		\ead{francois.fillion@emt.inrs.ca}
		
		\address[1]{Arctan, Inc., Arlington, VA 22201, USA} 
		\address[2]{Max Planck Institute for Nuclear Physics, 69117 Heidelberg, Germany}
		\address[3]{Tulane University, New Orleans, LA 70118, USA}
		\address[4]{Institute for Quantum Computing, University of Waterloo, Waterloo,
			Ontario N2L 3G1, Canada}
		\address[5]{Infinite Potential Laboratories, Waterloo, Ontario, Canada, N2L 0A9}
		\address[6]{Universit\'{e} du Qu\'{e}bec, INRS-\'{E}nergie, Mat\'{e}riaux et T\'{e}l\'{e}communications, Varennes, Qu\'{e}bec, Canada J3X 1S2}
		\cortext[cor1]{Corresponding author}

		\begin{abstract}
			A class of explicit numerical schemes is developed to solve for the relativistic dynamics and spin of particles in electromagnetic fields, using the Lorentz-BMT equation formulated in the Clifford algebra representation of Baylis. It is demonstrated that these numerical methods, reminiscent of the leapfrog and Verlet methods, share a number of important properties: they are energy-conserving, volume-conserving and second order convergent. These properties are analysed empirically by benchmarking against known analytical solutions in constant uniform electrodynamic fields. It is demonstrated that the numerical error in a constant magnetic field remains bounded for long time simulations in contrast to the Boris pusher, whose angular error increases linearly with time. Finally, the intricate spin dynamics of a particle is investigated in a plane wave field configuration. 
		\end{abstract}
		
		
	\end{frontmatter}
	

\section{Introduction}

The classical relativistic dynamics of charged particles in electromagnetic fields is ubiquitous in nature and as a consequence, is an important topic in many areas of physics, ranging from plasma physics, astrophysics, accelerator physics and many others \cite{jackson2007classical, PhysRevLett.43.267, blandford1978particle, spurio2014particles, https://doi.org/10.1029/2005GL024256}. In plasma physics, the main theoretical tools based on numerical simulations of the Vlasov equation, the so-called particle-in-cells (PIC) method, rely on accurate long term approximations of particle trajectories \cite{birdsall2004plasma, RevModPhys.55.403,https://doi.org/10.1002/ctpp.200710072, Arber_2015}. In particle accelerators, these trajectories are important to determine the stability of the beam in the storage ring, again requiring long time solutions \cite{chao2013handbook,wiedemann2015particle,lee2018accelerator}. 

Recently, some studies have pointed out the importance of spin dynamics in plasmas \cite{PhysRevLett.98.025001}. Also, it is well-known in particle physics that spin dynamics is important to prepare the electron beam in the right polarization, requiring fine tuning and control of the electromagnetic fields in the accelerator \cite{conte2008introduction,Mane_2005}.

Describing these physical systems theoretically then require two main ingredients: an equation that describes the (classical) state of the particle (position, velocity and spin) and an accurate approximation for the solution of this equation. The former is provided by the combination of the relativistic Lorentz equation, for charged particle trajectories, and the Bargmann-Michel-Telegdi (BMT) equation \cite{PhysRevLett.2.435}, which gives a classical description of spin precession when gradients of the field can be neglected \cite{jackson2007classical}. The latter is the subject of this article, where numerical methods are developed to solve these equations. 

Solving the Lorentz-BMT set of equations numerically and efficiently for long time simulations is a long-standing problem. To reach this goal, many numerical approaches have been developed over the years, most of them focusing on the Lorentz equation and neglecting spin. The quintessential numerical scheme is the Boris pusher \cite{boris1970relativistic}, developed in the 70's and now widely used in PIC codes and for simulating the particle dynamics in magnetic fields. Its success stems from the fact that the algorithm is simple and preserves the phase-space volume in the non-relativistic limit, despite not being symplectic \cite{doi:10.1063/1.4818428}. Moreover, it also preserves energy explicitly for certain field configurations \cite{hairer2018energy}. These properties make this method much more accurate in long time simulations than standard approaches for systems of ordinary differential equations, such as the Runge-Kutta methods, while still being easy to implement. Given its success, the Boris method has been revisited many times. For example, more accurate versions of the Boris method have been developed via a modified gyration angle update \cite{UMEDA20181,UMEDA201937} or a filter algorithm \cite{hairer2020filtered}. Also, a Boris-like algorithm with spatial stepping also exists \cite{PhysRevSTAB.5.094001}. Other alternatives to the Boris method in the non-relativistic limit includes high order exponential operator splitting \cite{PhysRevE.77.066401, PhysRevE.102.043315, HE2015135,wu2020explicit, WANG2019112617}, symplectic methods \cite{TAO2016245,WEBB2014570}, multisteps methods \cite{SMAI-JCM_2017__3__205_0} and the line integral method \cite{BRUGNANO2019209}.

In the relativistic regime, the system of equation becomes nonlinear and thus, more challenging to solve. In addition, the Boris method is no longer volume-conserving and thus, looses its accuracy in strong electromagnetic fields. For these reasons, many alternatives to the Boris approach have been developed recently to tackle the relativistic dynamics of charged particles \cite{doi:10.1063/1.2837054,doi:10.1063/1.4979989,doi:10.1063/1.4962677,doi:10.1063/1.4916570,petri_2017}. Most of these approaches rely on the explicit conservation of the phase space volume and/or the conservation of energy for long time accuracy. Many of the most popular numerical schemes are compared in Ref. \cite{Ripperda_2018}. Attempts to solve the BMT equation in conjunction with the Lorentz force equation are more rare however, but can be found in Refs. \cite{MEOT1992492,PhysRevSTAB.18.024001}.

In this article, we put forth new frameworks for simulating relativistic dynamics of trajectory and spin for charged particles in a strong electromagnetic field based on the spinor formulation of Baylis \cite{doi:10.1063/1.528135,Baylis_1989,baylis1992classical,BaylisBook1996,baylis2004electrodynamics}. Leapfrog-like and Verlet-like  second-order numerical methods are developed based on the operator splitting. Both   preserve the phase-space volume; additionally, energy is conserved when  electric field is absent. These properties are tested empirically by comparing computed  trajectories to known analytical solutions in homogeneous constant electric and magnetic fields. It is demonstrated that the numerical error stays bounded for all field configurations, in contrast to the Boris method, for which the numerical error increases linearly with time in the case of a strong constant magnetic field. Furthermore, our formulation provides a simple access to spin, which can be evaluated without solving another differential equation. The spin dynamics is benchmarked with the exact solution for a particle moving in a plane electromagnetic wave.

This article is organized as follow. In Section \ref{clifford}, we review the formalism of the Clifford algebras applied to electrodynamics. In Section \ref{sec:num_meth} the numerical
methods are described. In Section \ref{sec:num_res} the methods are benchmarked against the relativistic Boris method. We then close in Section \ref{Conclusion} with conclusions and an outlook.
Units where $c=1$ are used throughout this article.

\section{Review of the formalism of Clifford (Geometric) Algebras applied to electrodynamics \label{clifford}}
Throughout this paper, a classical particle 
of charge $q$ and mass $m$ is considered. Bold letters correspond to ordinary three-dimensional vectors.

\subsection{Electron dynamics}
In the usual classical relativistic formulation, the trajectory of a charged particle $\boldsymbol{x}$ is governed by the Lorentz-force equation
\begin{align} \label{Lorentz-Force-Eq}
 \frac{d \boldsymbol{p} }{dt} = q \left(\boldsymbol{E} +  \frac{d\boldsymbol{x}}{dt} \times \boldsymbol{B} \right),
\end{align}
where $\boldsymbol{p} = \gamma m d\boldsymbol{x}/dt$ denotes the momentum,\\ 
 $\gamma = 1/\sqrt{1-(d\boldsymbol{x}/dt)^2}$ is the Lorentz factor,
$\boldsymbol{E}$ and $\boldsymbol{B}$ are the electric and magnetic fields, respectively.
The  manifestly-covariant form of this equation reads
\begin{align} \label{Lorentz-Force-Cov-Eq}
 m \frac{d u^{\mu} }{d\tau} = q F^{\mu \nu} u_{\nu},
\end{align}
where   $u^{\mu} = (u^0,\boldsymbol{u})=\gamma(1,  d\boldsymbol{x}/dt )$ is the proper velocity,
 $\tau$ is the proper time, and $F^{\mu\nu}$ is the electromagnetic field tensor.
The Einstein summation convention is assumed over repeated   
Greek indices $\mu,\nu=0,1,2,3$.   

In the Clifford algebra formalism, the motion and orientation of a particle is determined by its eigenspinor $\Lambda$, which
is just the special Lorentz transformation relating the rest frame of the charge to the lab frame.
The properties of spacetime vectors known in the rest frame of the charge are transformed to the
lab frame by $\Lambda$. For instance, the proper velocity of the particle in the lab frame is
\begin{align}\label{properVelSpinor}
U=\Lambda\Lambda^\dagger,
\end{align}
where
\begin{align}
 \label{definition-U}
 U &  = (\sigma_0u^0+\sigma_1u^1+\sigma_2u^2+\sigma_3u^3)\nonumber\\
 &=\begin{pmatrix}
       u^0 + u^3 & u^1 - i u^2 \\ u^1 + i u^2 & u^0 - u^3
      \end{pmatrix}.
\end{align}
The component $\sigma_{0}$ is the $2\times2$ identity matrix while $\sigma_k$, $k=1,2,3$, are the Pauli matrices. 
Note that both matrices $U$ and $\Lambda$ are unimodular 
owing to the mass-shell condition  $\det(U) =  (u^0)^2 - \boldsymbol{u}^{2} =1$. The velocity in the usual quadri-vector representation can be recovered via
\begin{align}
 u^{\mu} = \frac{1}{2}\Tr(U \sigma_{\mu}).
\end{align}


As shown in \cite{hestenes1974proper,BaylisClassicalSpinorEMW1999,baylis2004electrodynamics}, the 
Lorentz-force equation (\ref{Lorentz-Force-Cov-Eq}) 
can be  written in terms of $\Lambda$ as
\begin{align}
 \label{spinorial-LF}
 \frac{d \Lambda}{d \tau} = \frac{q}{2m } F \Lambda,
\end{align}
where $F$ is the electromagnetic field tensor represented by the traceless matrix  
\begin{align} \label{F-field}
 F &= E^k\sigma_k+iB^k\sigma_k\nonumber\\
 &=\begin{pmatrix}
       E^3 & E^1 - i E^2 \\ E^1 + i E^2 &  - E^3
      \end{pmatrix} + i
      \begin{pmatrix}
       B^3 & B^1 - i B^2 \\ B^1 + i B^2 &  - B^3
      \end{pmatrix}.\nonumber\\
\end{align}
The equivalence between Eqs. (\ref{Lorentz-Force-Cov-Eq}) and (\ref{spinorial-LF}) can be proven as follows. Taking the proper time derivative of equation (\ref{properVelSpinor}) gives
\begin{align}
\frac{dU}{d\tau}=\frac{d\Lambda}{d\tau}\Lambda^\dagger+\Lambda\frac{d\Lambda^\dagger}{d\tau}.
\end{align}
It follows from Eq. (\ref{spinorial-LF}) that
\begin{align}
\frac{d\Lambda}{d\tau}\Lambda^\dagger=\frac{q}{2m}FU,\quad \Lambda \frac{d\Lambda^\dagger}{d\tau}=\frac{q}{2m}UF^\dagger.
\end{align}
Thus,
\begin{align}
\frac{dU}{d\tau}&=\frac{q}{2m}(FU+UF^\dagger) ,\nonumber\\
&=\frac{q}{m}\left(\sigma_0\boldsymbol{E}\cdot\boldsymbol{u}+u^0E^k\sigma_k+\sigma_k(\boldsymbol{u}\times \boldsymbol{B})^k\right),
\end{align}
given that $U$ is hermitian. Therefore, the scalar and vector parts of the above equation give, respectively
\begin{align}
m\frac{du^0}{d\tau}&=q\boldsymbol{E}\cdot\boldsymbol{u},\label{zeroth-Lorentz}\\
m\frac{d\boldsymbol{u}}{d\tau}&=u^0 q \boldsymbol{E}+(\boldsymbol{u}\times q\boldsymbol{B}).\label{LorentzV}
\end{align}
Since the eigenspinor is related to the 
proper velocity, the spacetime trajectory $x^{\mu}$ is recovered as   
\begin{align} 
\label{proper-velocity-general}
 x^{\mu} &=  \frac{1}{2}\Tr( x \sigma_{\mu}  ), \\
 \frac{dx}{d \tau } &=  \Lambda \Lambda^{\dagger} \label{proper_velocity-eq} . 
\end{align}  

The eigenspinor of a particle is different for different observers. Suppose that $\Lambda_{A}$ is the eigenspinor of a charged particle with respect to observer $A$. Let $L_{BA}$ transform properties from the rest frame of the observer $A$ as viewed by observer $B$. The eigenspinor for observer $B$ is then
$$
\Lambda_B=L_{BA}\Lambda_A.
$$
The transformation of the eigenspinor thus takes the form
$$
\Lambda\rightarrow L\Lambda.
$$

\subsection{Spin dynamics}
The most significant advantage of the spinorial propagator 
is the ability to provide the classical spin dynamics as described 
by the BMT equation. Arbitrarily defining  $\sigma_3$ as the direction of the spin in the particle's rest frame, the spin 4-vector 
in the laboratory frame is then given by  \cite{Baylis2004-BMT}
\begin{align}\label{SpinDef}
 S = \Lambda\sigma_3 \Lambda^\dagger, 
\end{align}
where $\Lambda$ obeys the dynamical equation Eq. (\ref{spinorial-LF})
for the $g$-factor $g=2$. 
Taking the proper time derivative of \eqref{SpinDef}, we have
\begin{align}
\frac{dS}{d\tau}=\frac{d\Lambda}{d\tau}\sigma_3\Lambda^\dagger+\Lambda\sigma_3\frac{d\Lambda^\dagger}{d\tau}.
\end{align}
It follows from Eq. (\ref{spinorial-LF}) that
\begin{align}
\frac{d\Lambda}{d\tau}\sigma_3\Lambda^\dagger=\frac{q}{2m}FS,\quad \Lambda \frac{d\Lambda^\dagger}{d\tau}=\frac{q}{2m}SF^\dagger.
\end{align}
Thus,
\begin{align}
\frac{dS}{d\tau}&=\frac{q}{2m}(FS+SF^\dagger) \nonumber\\
&=\frac{q}{m}\left(\sigma_0\boldsymbol{E}\cdot\boldsymbol{S}+S_0E^k\sigma_k+\sigma_k(\boldsymbol{S}\times \boldsymbol{B})^k\right)
\end{align}
given that $U$ is hermitian. Therefore, collecting the terms and writing in covariant form, we end up with the
BMT equation
\begin{align}
\frac{dS^\alpha}{d\tau}=\frac{q}{m}F^{\alpha\beta}S_\beta.
\end{align}
In the standard approach, this differential equation is solved along with the Lorentz-force equation \eqref{Lorentz-Force-Cov-Eq}. This is a challenging problem because the two equations are coupled via the electromagnetic field. In the Clifford algebra formulation, we solve for $\Lambda$ by using Eq. \eqref{spinorial-LF} and the spin is simply evaluated using Eq. \eqref{SpinDef}.

\section{Numerical methods \label{sec:num_meth}}

In this section, a class of numerical schemes is developed starting from the Lorentz-BMT force equation formulated in the Clifford algebra of Baylis. The main physical goal is obtaining accurate relativistic trajectories of particles immersed in a space-time dependent electromagnetic field. This will be achieved by a combined use of the split-operator method and standard discretization techniques, resulting in simple but efficient numerical methods.  

The starting point is the system of ordinary differential equation (ODE) obeyed by the particle in its proper reference frame, obtained from Eq. \eqref{properVelSpinor}, along with Eq. \eqref{spinorial-LF}. The proper time is related to the lab frame time by $dt /d\tau = \gamma(t)$, where $\gamma(t)$ is the Lorentz factor. Then, the particle dynamical equations become
\begin{align}
\label{eq:Lambda_lab}
\cfrac{d\Lambda(x(t))}{dt} &= \cfrac{q}{2m \gamma (t)} F(x(t)) \Lambda(x(t)), \\
\label{eq:x_lab}
\cfrac{dx(t)}{dt} &= \frac{1}{\gamma(t)} \Lambda(x(t)) \Lambda^{\dagger}(x(t)), 
\end{align}
where $\Lambda(x(t)) \in \mathrm{M}_{2}(\mathbb{C})$ is the eigenspinor describing the motion and orientation of a particle, $x(t) \in \mathrm{M}_{2}(\mathbb{C})$ is the position of the particle and $F \in \mathrm{M}_{2}(\mathbb{C})$ is the electromagnetic tensor. All these quantities are expressed in the Clifford algebra described in Sec. \ref{clifford}, where a Pauli matrix basis decomposition is given [see Eqs. \eqref{F-field} and \eqref{proper_velocity-eq}].

Together with the initial values $x(t_{0}) = x^{0}$ and $\Lambda(x^{0}) = \Lambda^{0}$, Eqs. \eqref{eq:Lambda_lab}-\eqref{eq:x_lab} form the initial value problem solved by the numerical methods. The initial position $x^{0}$ is evaluated from Eq. \eqref{proper_velocity-eq}. $\Lambda^{0}$ can be obtained by using the fact that $\Lambda$ is a unimodular element of the Pauli algebra, and therefore can be written as a pure boost \cite{doi:10.1063/1.528135,Baylis_1989,doi:10.1119/1.17736}:
\begin{align}
\Lambda^{0} =  e^{\frac{w^{0}}{2}},
\end{align}
where $w^{0}$ is the initial rapidity. The rapidity is given by
\begin{align}
w^{0} = \boldsymbol{\sigma} \cdot \hat{\boldsymbol{u}}^{0} \arctan(|\boldsymbol{u}^{0}|),
\end{align}
where $\hat{\boldsymbol{u}}^{0}:= \boldsymbol{u}^{0}/|\boldsymbol{u}^{0}|$ is the unit vector in the direction of the initial velocity $\boldsymbol{u}^{0} \in \mathbb{R}^{3}$.

The ODE system \eqref{eq:Lambda_lab} and \eqref{eq:x_lab} has an important mathematical property: it preserves the phase space volume. As presented in \ref{app:vol_ode}, this can be demonstrated by showing that the ODE is divergenceless \cite{quispel2001six}. It is emphasized here that the phase space is spanned by the position and eigenspinor $(x,\Lambda)$. In particular, it is not the same space as the one for Hamiltonian systems, defined via the position and momentum of the particle $(\boldsymbol{x},\boldsymbol{p})$. Nevertheless, volume preservation is an intrinsic property of the dynamic ODE system \eqref{eq:Lambda_lab}-\eqref{eq:x_lab}  and therefore, numerical schemes fulfilling this property should be more accurate in long term calculations because they will preserve the qualitative features of the solution \cite{iserles2016geometric}.

To develop such numerical schemes, a time grid is introduced where $x^{n} = x(t_{n})$, $\Lambda^{n} = \Lambda(x(t_{n}))$, and $t_{n} = t_{0} + n \Delta t$, where $n \in \mathbb{N}$ and $\Delta t$ is the time step. To preserve volume, the two approaches described in the following subsections take advantage of the fact that the formal solution of Eq. \eqref{eq:Lambda_lab} is 
\begin{align}
\Lambda^{n+1} &= T\exp \left[  \cfrac{q}{2m} \int_{t_{n}}^{t_{n+1}} \frac{F(x(t') )}{\gamma(t')} dt ' \right] \Lambda^{n}, 
\end{align}
where $T$ represents time-ordering. This can be written in a form more convenient for numerical approximation \cite{Suzuki1993}:
\begin{align}
\Lambda^{n+1} &= \exp \left[\Delta t \left(\frac{q}{2m}\frac{F(x^{n})}{\gamma^{n}} + \mathcal{T}\right) \right] \Lambda^{n},
\end{align}
where $\mathcal{T} =  \overleftarrow{\partial_{t_{n}}}$ is now the ``left'' time-shifting operator. At this point, the solution is still exact. To evaluate this numerically, an operator splitting approximation scheme is implemented. In particular, a third order accurate approximation of the last expression,  the symmetric exponential decomposition, is used. It is given by
\begin{align}
\Lambda^{n+1}
&=e^{\frac{\Delta t}{2}  \mathcal{T}}
\exp \left[\Delta t \frac{q}{2m}\frac{F(x^{n})}{\gamma^{n}} \right]
e^{\frac{\Delta t}{2} \mathcal{T}}\Lambda^{n} + O(\Delta t^{3}), \\
\label{eq:lambda_scheme}
&= \exp \left[\Delta t \frac{q}{2m}\frac{F(x^{n+\frac{1}{2}})}{\gamma^{n+\frac{1}{2}}} \right] \Lambda^{n}+ O(\Delta t^{3}) , \\
\label{eq:lambda_scheme_u}
&= U^{n} \Lambda^{n}
\end{align} 
where the properties of the time-shifting operator has been used to obtain \eqref{eq:lambda_scheme} and where $U^{n}$ is the two-by-two transition matrix. The latter corresponds to an exponential scheme for the $\Lambda$-update, typical of operator splitting methods.  It allows for estimating $\Lambda^{n+1}$ assuming $\Lambda^{n}$, $x^{n+\frac{1}{2}}$ and $\gamma^{n+\frac{1}{2}}$ are known.  

However, the Lorentz factor is related to $\Lambda$ via \eqref{proper_velocity-eq} and therefore, is available only at time $t_{n}$.  An accurate approximation of the Lorentz factor at $t_{n+\frac{1}{2}}$ can be obtained by deriving an equation for its time-dependence and by approximating this evolution equation to a desired order. Taking the time derivative of $\gamma(t)=\mathrm{Tr}[\Lambda\Lambda^\dagger]/2$ gives
\begin{align}
\frac{d\gamma(t)}{dt} &= \frac{1}{2} \mathrm{Tr} \left[ \frac{d\Lambda}{dt} \Lambda^{\dagger} + \Lambda \frac{\Lambda^{\dagger}}{dt}\right], \notag\\
&= \frac{q}{4m\gamma(t)}  \mathrm{Tr} \left[ F \Lambda \Lambda^{\dagger} + \Lambda \Lambda^{\dagger} F^{\dagger} \right],
\end{align}
where Eq. \eqref{eq:Lambda_lab} was used to get the second equation. The latter can be discretized to obtain the value of the Lorentz factor at $t_{n+\frac{1}{2}}$ with a second order accuracy, in order to be consistent with the accuracy of the exponential evolution scheme for $\Lambda$ in Eq. \eqref{eq:lambda_scheme}. An explicit Euler method is used for that purpose, yielding
\begin{align}
\label{eq:gamma_update}
\gamma^{n+\frac{1}{2}} &= \gamma^{n} + \frac{\Delta t}{2} \frac{q}{4m\gamma^{n}}  \mathrm{Tr} \left[ F(x^{n}) \Lambda^{n} \Lambda^{n\dagger} + \Lambda^{n} \Lambda^{n\dagger} F^{\dagger}(x^{n}) \right] \nonumber \\
&+ O(\Delta t^{2}) .
\end{align}
When this second order accurate expression is reported into Eq. \eqref{eq:lambda_scheme}, it incurs a third order error on the exponential, consistent with the numerical scheme. 

The last ingredient missing for the update of $\gamma$ and $\Lambda$ is the position $x$, evaluated at times $t_{n}$ and $t_{n+\frac{1}{2}}$. This can be achieved by evolving $x$ on half time-steps or on a time-staggered grid, in the same spirit as the Verlet and leap-frog methods, respectively. This will be described in more detail in the following subsections.

\subsection{Verlet-like numerical scheme \label{sec:verlet}}

A Verlet-like numerical scheme is obtained by approximating Eq. \eqref{eq:x_lab} using a two-step method, based on the explicit forward and backward Euler scheme:
\begin{align}
\label{eq:verlet_x_update1}
x^{n+\frac{1}{2}} &= x^{n} + \frac{\Delta t}{2}  \frac{\Lambda^{n} \Lambda^{n \dagger}}{\gamma^{n}} + O(\Delta t^{2}), \\
\label{eq:verlet_x_update2}
x^{n+1} &= x^{n+\frac{1}{2}} + \frac{\Delta t}{2}  \frac{\Lambda^{n+1} \Lambda^{n+1 \dagger}}{\gamma^{n+1}} + O(\Delta t^{2}). 
\end{align}
Although each step has an accuracy $O(\Delta t^{2})$, the full evolution is $O(\Delta t^{3})$. This can be demonstrated by substituting Eq. \eqref{eq:verlet_x_update1} into \eqref{eq:verlet_x_update2}. Then, we get
\begin{align}
x^{n+1} &= x^{n} + \frac{\Delta t}{2}  \left[ \frac{\Lambda^{n} \Lambda^{n \dagger}}{\gamma^{n}}  +   \frac{\Lambda^{n+1} \Lambda^{n+1 \dagger}}{\gamma^{n+1}} \right], 
\end{align}
corresponding to the trapezoidal rule method with an accuracy $O(\Delta t^{3})$. Splitting this in two steps as in Eqs. \eqref{eq:verlet_x_update1}-\eqref{eq:verlet_x_update2} allows for getting the position at time $t_{n+\frac{1}{2}}$ required in Eq. \eqref{eq:lambda_scheme}. 

To summarize, here is a description of the algorithm to evolve the position for one time step. It assumes that $\Lambda^{n}$, $x^{n}$ are known: 
\begin{enumerate}
	\item Compute $\gamma^{n}$ using  $\gamma^{n} = \frac{1}{2} \mathrm{Tr}\left[\Lambda^{n}\Lambda^{n\dagger}\right]$.
	\item Compute $x^{n+\frac{1}{2}}$ using Eq. \eqref{eq:verlet_x_update1}.
	\item Compute $F(x^{n})$. 
	\item Compute $\gamma^{n+\frac{1}{2}}$ using Eq. \eqref{eq:gamma_update}.
	\item Compute $\Lambda^{n+1}$ using Eq. \eqref{eq:lambda_scheme}. 
	\item Compute $\gamma^{n+1}$ using  $\gamma^{n+1} = \frac{1}{2} \mathrm{Tr}\left[\Lambda^{n+1}\Lambda^{n+1 \dagger}\right]$. 
	\item Compute $x^{n+1}$ using Eq. \eqref{eq:verlet_x_update2}.
\end{enumerate}

\subsection{Leapfrog-like numerical scheme \label{sec:leapfrog}}

The leapfrog-like scheme is obtained by considering a time-staggered grid, where $\Lambda$ and $x$ are evaluated on different time steps. Then, the time derivative in Eq. \eqref{eq:x_lab} is discretized using a midpoint finite difference scheme, centered on $t_{n}$. This is written as
\begin{align}
\label{eq:x_update}
x^{n+\frac{1}{2}} = x^{n-\frac{1}{2}} + \Delta t  \frac{\Lambda^{n} \Lambda^{n \dagger}}{\gamma^{n}} + O(\Delta t^{3}).
\end{align}  
This again has an accuracy $O(\Delta t^{3})$. 

With this staggered grid, the position is not evaluated at $t_{n}$, as required to obtain the Lorentz factor at $t_{n+\frac{1}{2}}$. The strategy used here is to    
approximate the electromagnetic field in Eq. \eqref{eq:gamma_update} by linear interpolation as
\begin{align}
\label{eq:F_ave}
 F(x^{n}) = \frac{ F(x^{n+\frac{1}{2}}) +  F(x^{n-\frac{1}{2}})}{2} + O(\Delta t^{2}).
\end{align}
This average can be evaluated on the staggered grid.

To summarize, here is a description of the algorithm for one time step. It assumes that $\Lambda^{n}$, $x^{n-\frac{1}{2}}$ are known: 
\begin{enumerate}
	\item Compute $\gamma^{n}$ using $\gamma^{n} = \frac{1}{2} \mathrm{Tr}\left[\Lambda^{n}\Lambda^{n\dagger}\right]$.
	\item Compute $x^{n+\frac{1}{2}}$ using Eq. \eqref{eq:x_update}.
	\item Compute $F(x^{n-\frac{1}{2}})$ and $F(x^{n+\frac{1}{2}})$.
	\item Compute $F(x^{n})$ using Eq. \eqref{eq:F_ave}. 
	\item Compute $\gamma^{n+\frac{1}{2}}$ using Eq. \eqref{eq:gamma_update}.
	\item Compute $\Lambda^{n+1}$ using Eq. \eqref{eq:lambda_scheme}.
\end{enumerate}
The first step of the scheme, from $t_{0}$ to $t_{\frac{1}{2}}$, can be performed via the forward Euler step \eqref{eq:verlet_x_update1}. Although this step is $O(\Delta t^{2})$, it does not deteriorate the global convergence order of the numerical scheme because it is used only once.

\subsection{General properties of the numerical schemes \label{sec:gen_prop}}

The numerical schemes described in the last two subsections share a number of interesting properties. First, they have a second order rate of global convergence. Henceforth, the numerical error $\epsilon$ after $N$ time steps obeys 
\begin{align}
\label{eq:error}
\epsilon := \Vert x^{N} - x_{\mathrm{exact}}(t^{N}) \Vert \leq C \Delta t^{2},
\end{align}
where $C \in \mathbb{R}^{+}$ is some constant, $x^{N}$ is the approximated solution and $x_{\mathrm{exact}}$ is the exact solution. This statement is not proven rigorously here as this would demand a careful analysis of the regularity of the solution, which is outside the scope of this article. Rather, it is assumed that the solution is smooth enough, which is reasonable for a large class of physically relevant initial conditions and electromagnetic fields. In this case, the global convergence rate is usually one order less than the local accuracy. As demonstrated in Sections \ref{sec:verlet} and \ref{sec:leapfrog}, each step of the numerical schemes incurs local numerical error $O(\Delta t^{3})$, leading to a second order global rate of convergence. This property will be verified  empirically in Section \ref{sec:num_res}, where numerical results are displayed. 

Second, both numerical methods are energy conserving when there is no electric field $\boldsymbol{E} = 0$. This can be demonstrated in the following way. The energy $E^{N}$ of the particle is given after $N$ time steps by
\begin{align}
E^{N} = \gamma^{N} m = \frac{m}{2} \mathrm{Tr} \left[ \Lambda^{N} \Lambda^{N \dagger} \right].
\end{align}
On the other hand, from Eq. \eqref{eq:lambda_scheme_u}, we have that 
\begin{align}
\Lambda^{N} = \prod_{i=0}^{N} U^{i} \Lambda^{0}.
\end{align}
When the electric field is zero, the transition matrices $(U^{i})_{i=0,\cdots, N}$ are unitary, as can be deduced from the definition of $F$ in Eq. \eqref{F-field}. As a consequence, the energy becomes
\begin{align}
E^{N} = \frac{m}{2} \mathrm{Tr} \left[ \Lambda^{0} \Lambda^{0 \dagger} \right] = \gamma^{0} m,
\end{align} 
where the cyclic property of the trace and unitarity have been used to cancel the transition matrices. The fact that $E^{N} = E^{0}$ confirms that the energy is manifestly conserved by the numerical scheme. 

Finally, the third property of the numerical schemes is phase-space volume preservation. The detailed proof, given in \ref{app:vol_num_schemes}, hinges on the fact that the Jacobian of the flow has a unit determinant, for each step of the numerical schemes. As mentioned earlier, this property is important for long term simulations required in accelerator and plasma physics.

\section{Numerical results \label{sec:num_res}}

The Verlet-like and leapfrog-like numerical schemes have been implemented in C++, using the highly efficient and easy to use linear algebra library A\textsc{rmadillo} \cite{Sanderson2016}. The resulting code can perform approximately $5.0 \times 10^{4}$ time steps per second on a standard laptop computer (with an Intel I7 CPU). The numerical methods are compared to the standard Boris pusher, described in Ref. \cite{Ripperda_2018} and implemented in Python. To verify some numerical properties and benchmark against known analytical solutions, simple uniform electromagnetic fields are first considered, in the same spirit as the numerical tests given in Ref. \cite{Ripperda_2018}. Then, to display nontrivial spin dynamics, a plane wave electromagnetic field is chosen.

\subsection{Constant uniform electric field}

A uniform electric field applies a force on a charged particle, inducing acceleration in the field orientation. Without loss of generality, we consider an electric field pointing in the $x$-coordinate given by $\boldsymbol{E} = (E,0,0)$. In this simple case, the Lorentz equation of motion can be solved analytically. Assuming the particle is initially at rest ($\boldsymbol{v}(0) = 0$) and positioned at the origin ($\boldsymbol{x}(0) = 0$), the solution is given by \cite{Ripperda_2018}
\begin{align}
x_{\mathrm{analytical}}(t) &= \frac{m}{qE} \left[ \gamma(t) - 1 \right]
\end{align} 
where 
\begin{align}
\gamma(t) &= \sqrt{1+ \frac{(qEt)^{2}}{m}}.
\end{align}

\begin{figure}[t]
	\begin{center}
		\includegraphics[width=0.5\textwidth]{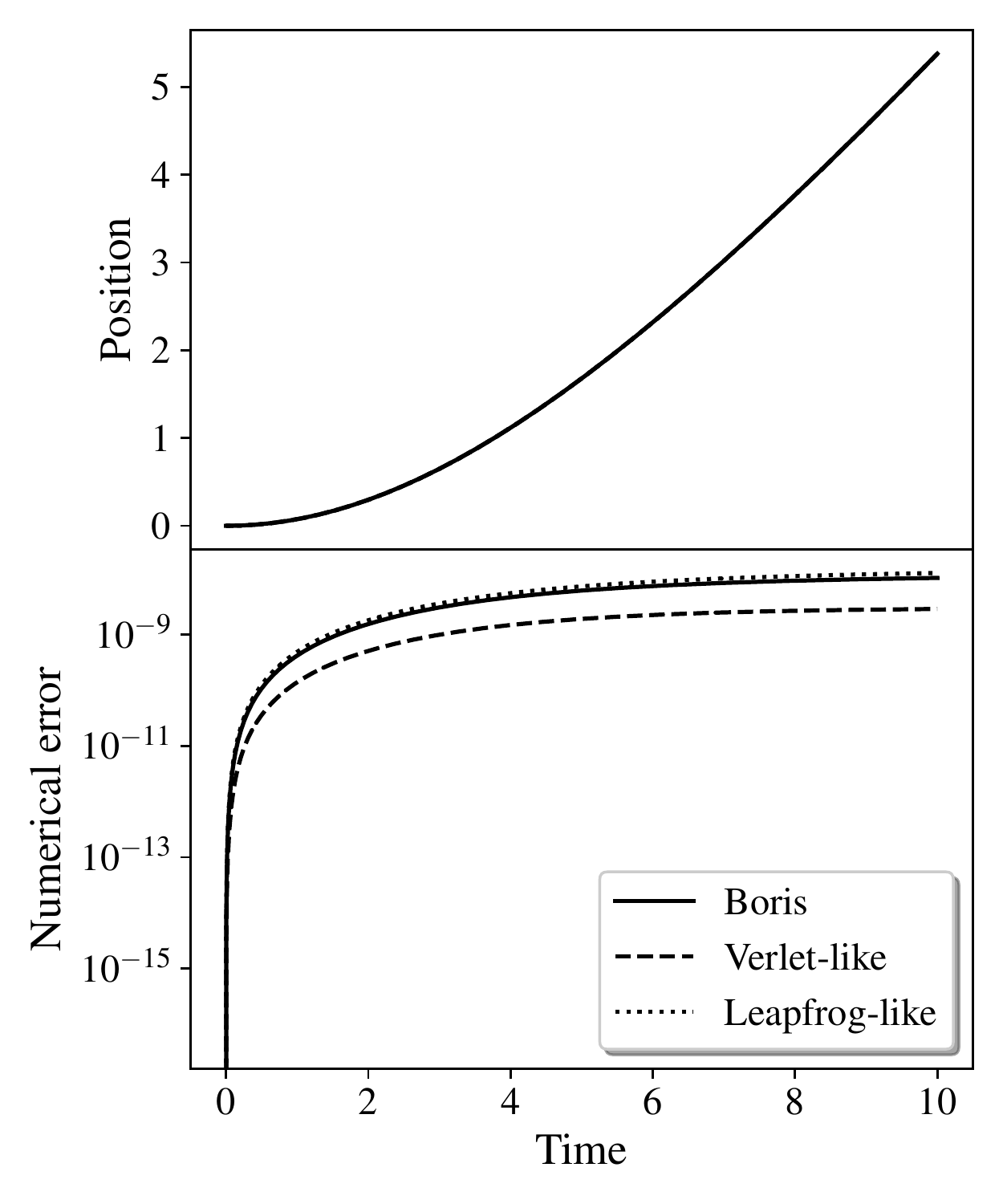} 
	\end{center}
	\caption{Position $x$ and numerical error $\epsilon(t) = |x(t)-x_{\mathrm{analytical}}(t)|$ as function of time for the Boris, Verlet-like and Leapfrog-like methods. All position curves are overlapping with the analytical solution. }
	\label{fig:num_error_t}
\end{figure}

To test the numerical methods, we consider a positively charged particle with an electron mass (a positron with mass $m=1$ in natural units) immersed in an electric field of magnitude $E=0.5$. In the first test, we look at the particle position as a function of time and compare to the analytical solution. The final time of the simulation is set to $t_{\mathrm{final}} = 10.$ and the number of time steps to $N=10000$, making for a time step of $\Delta t = 1.0 \times 10^{-3}$. The numerical results for the position are displayed in Fig. \ref{fig:num_error_t}, along with the numerical error $\epsilon$ evaluated from Eq. \eqref{eq:error}. The results demonstrate that all methods reproduce accurately the analytical solution (all the curves are overlapping). However, the error of the Verlet-like scheme is lower than the two other methods.  

\begin{figure}[t]
	\begin{center}
		\includegraphics[width=0.50\textwidth]{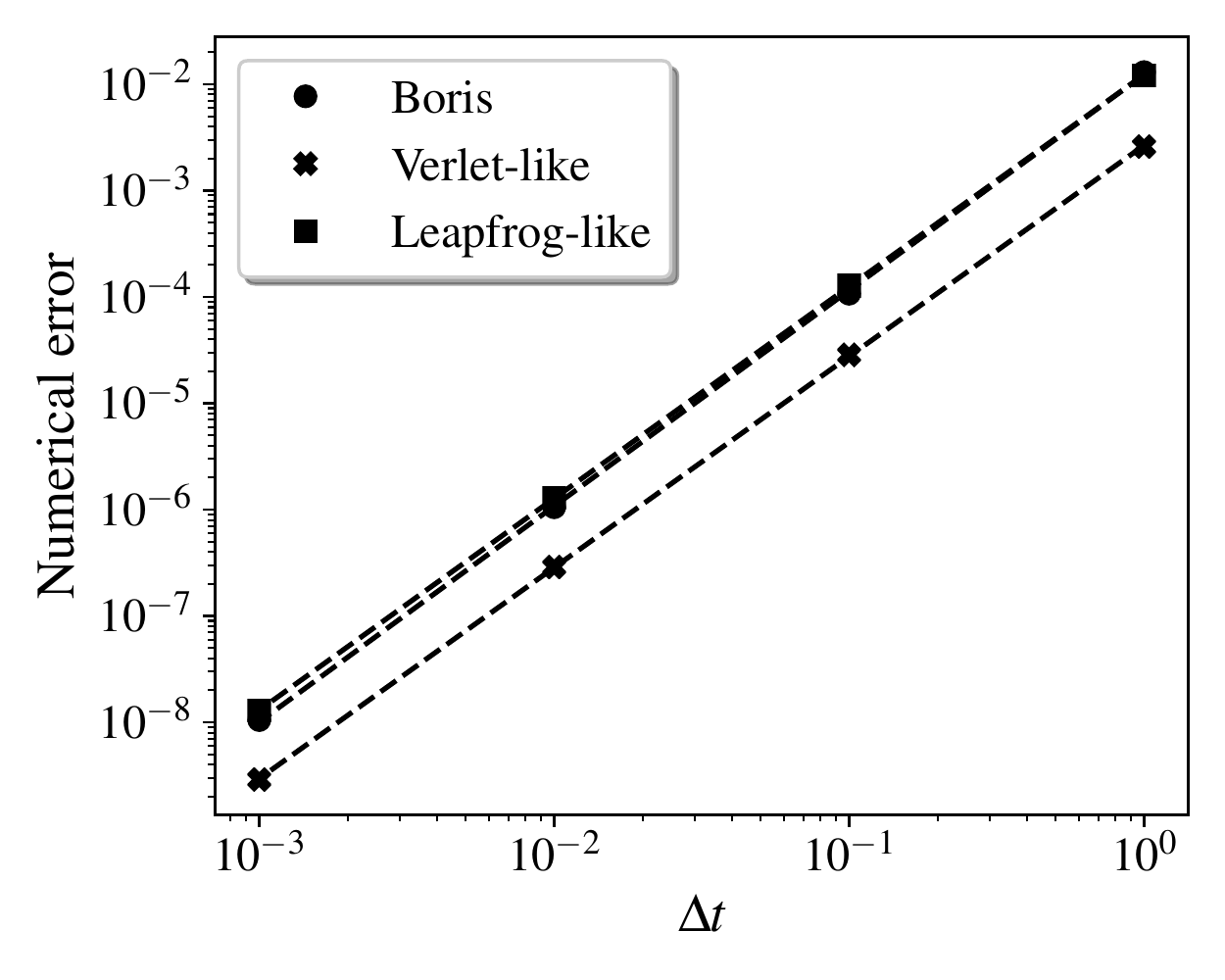} 
	\end{center}
	\caption{Numerical error at $t=t_{\mathrm{final}} = 10$  as function of the time step $\Delta t$ for the Boris, Verlet-like and Leapfrog-like methods. The dashed line corresponds to a fit of the data, used to determine the order of convergence. }
	\label{fig:num_error_dt}
\end{figure}

In the second test, we determine the order of convergence by looking at the scaling of the numerical error with the time step. The same particle, electric field and evolution time are considered. Four different number of time steps are chosen for each numerical methods, leading to different value of $\Delta t$. At the end of the simulation, the numerical error on position is evaluated using Eq. \ref{eq:error}. The numerical results are displayed in Fig. \ref{fig:num_error_dt}, along with the linear fit (dashed line) used to determine the order of convergence. The values of the order of convergence are given in Table \ref{tab:conv_order}. Similar to the first test, these numerical results demonstrate that numerical errors for the Boris and Leapfrog-like schemes are similar, while the Verlet-like method shows an improvement of approximately one order of magnitude, for any time step size. In addition, the analysis reveals that all the numerical schemes have a second-order convergence rate.

\begin{table}[t]
	\centering
	\begin{tabular}{lc}
		\hline \hline
		Numerical scheme & Order of convergence \\
		\hline
		Boris  &       2.0285\\
		Verlet-like &  1.9847 \\
		Leapfrog-like &  1.9916\\
		\hline \hline
	\end{tabular}
	\caption{Order of convergence for all the numerical schemes, determined from a fit of the error as a function of the time step.}
	\label{tab:conv_order}
\end{table}

\subsection{Constant uniform magnetic field}

In a constant magnetic field, a charged particle follows a circular trajectory at constant velocity $\boldsymbol{v}$ because the magnetic field do not exert any work on the particle. For simplicity, we choose a magnetic field in the $z$-coordinates, given by $\boldsymbol{B} = (0,0,B)$. In this case, the trajectory will follow a circle in the $xy$-plane. This is confirmed by looking that the analytical solution obtained from solving the Lorentz equation. The position is given by
\begin{align}
r = r_{g} \frac{\gamma_{v} m |\boldsymbol{v}|}{qB} \;\; , \;\;
\theta(t) = \frac{qB}{\gamma_{v} m} t,
\end{align}
where $\gamma_{v} = 1/\sqrt{1-\boldsymbol{v}^{2}}$ is the constant Lorentz factor, $r=\sqrt{x^{2} + y^{2}}$ is the radial distance from the origin, $r_{g}$ stands for the constant gyroradius and $\theta(t)$ is the angle with respect to the $y$-axis. Therefore, in simulations, the particle is positioned at $\boldsymbol{x} = (0,r_{g},0)$ at initial time $t=0$.

Again, we consider a positively charged particle with an electron mass. The magnitude of the magnetic field is set to $B=0.5$ while the initial velocity is chosen as $\boldsymbol{v} = (0.4,0.0,0.0)$. With these values, the gyroradius of the trajectory is $r_{g} \approx 2.8824564017956553$. 

\begin{figure}[t]
	\begin{center}
		\includegraphics[width=0.50\textwidth]{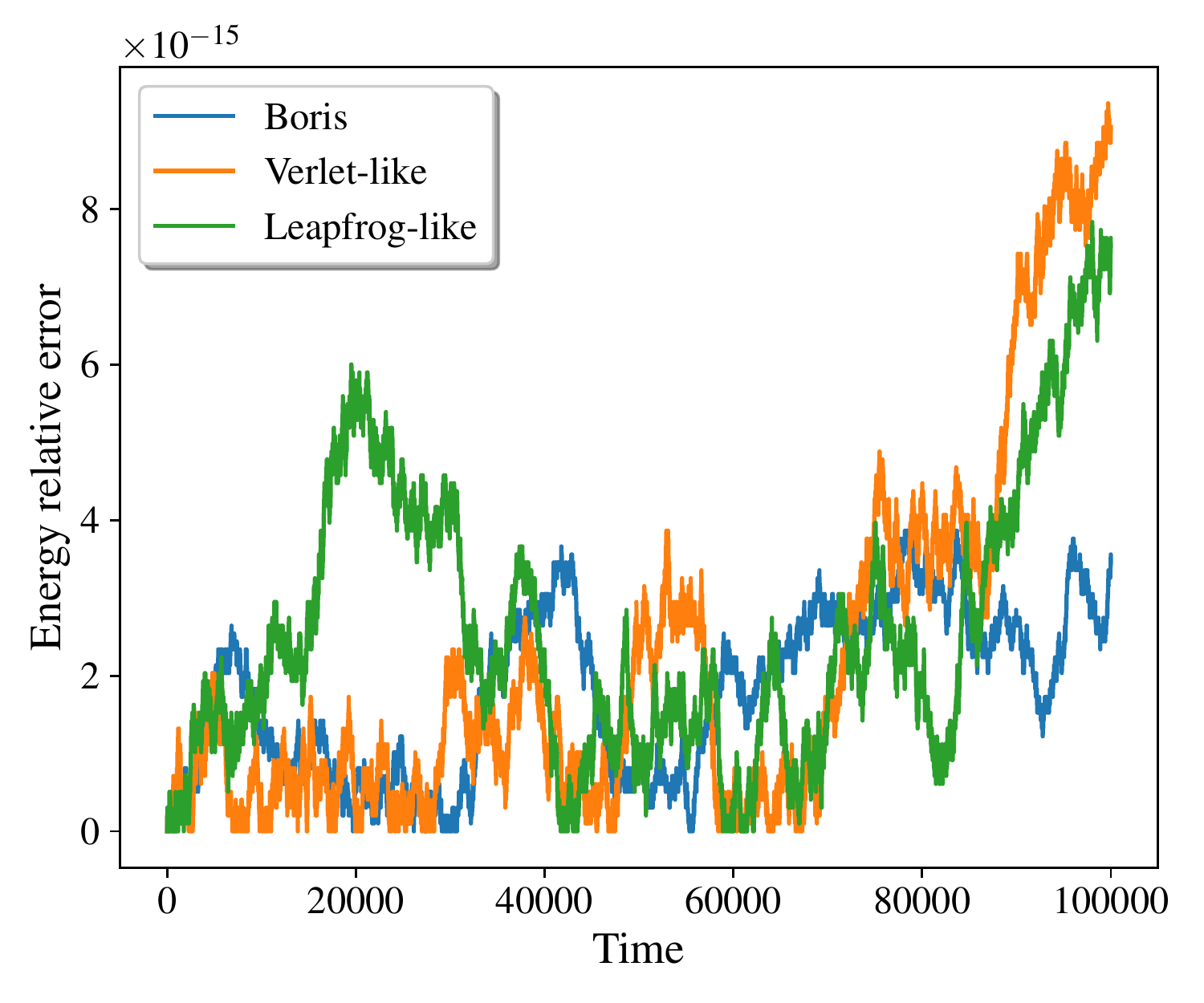} 
	\end{center}
	\caption{Relative error on energy as function of time. }
	\label{fig:energy_error}
\end{figure}

In the first test, we verify the conservation of energy claimed in Section \ref{sec:gen_prop}. To achieve this goal, a long term simulation is carried out with a final time of $t_{\mathrm{final}} = 100000$ and a number of time steps set to $N = 1.0 \times 10^{6}$. The energy is evaluated from the relativistic gamma factor as $E(t) = \gamma(t) m$. According to the exact solution, the energy is constant and given by $E_{\mathrm{exact}} = \gamma_{v} m$. In Fig. \ref{fig:energy_error}, the relative error on the energy, defined as $\epsilon_{\mathrm{rel}} = |E(t) - E_{\mathrm{exact}}|/ |E(t) + E_{\mathrm{exact}}|$, is displayed for the three numerical methods. They all show an excellent energy-preservation property, accurate up to machine precision. This result is an empirical confirmation of the theoretical result given in Section \ref{sec:gen_prop}.

\begin{figure*}[t]
	\begin{center}
		\includegraphics[width=1.0\textwidth]{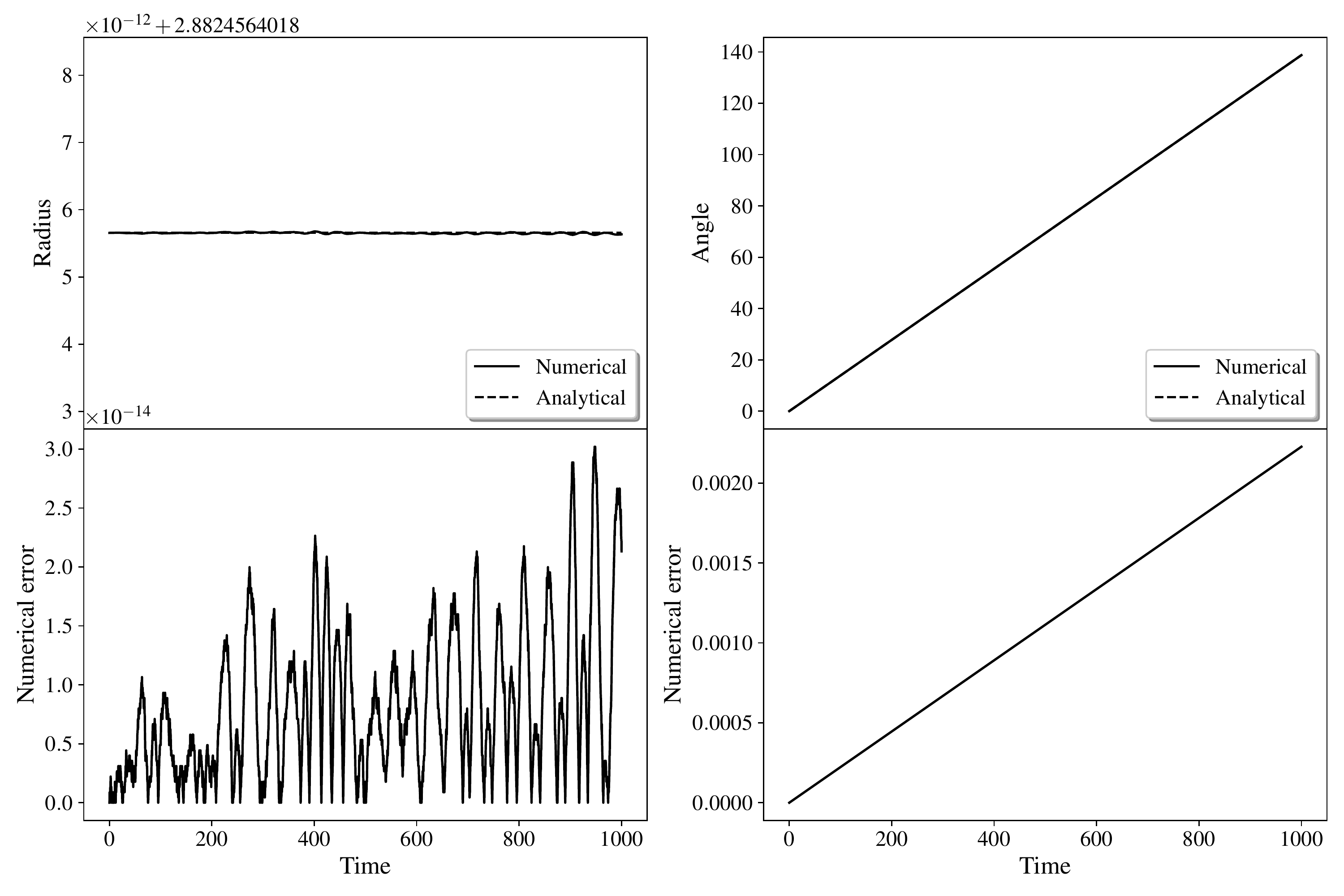} 
	\end{center}
	\caption{Relative error on energy as function of time. }
	\label{fig:consB_pos_boris}
\end{figure*}

\begin{figure*}[t]
	\begin{center}
		\includegraphics[width=1.0\textwidth]{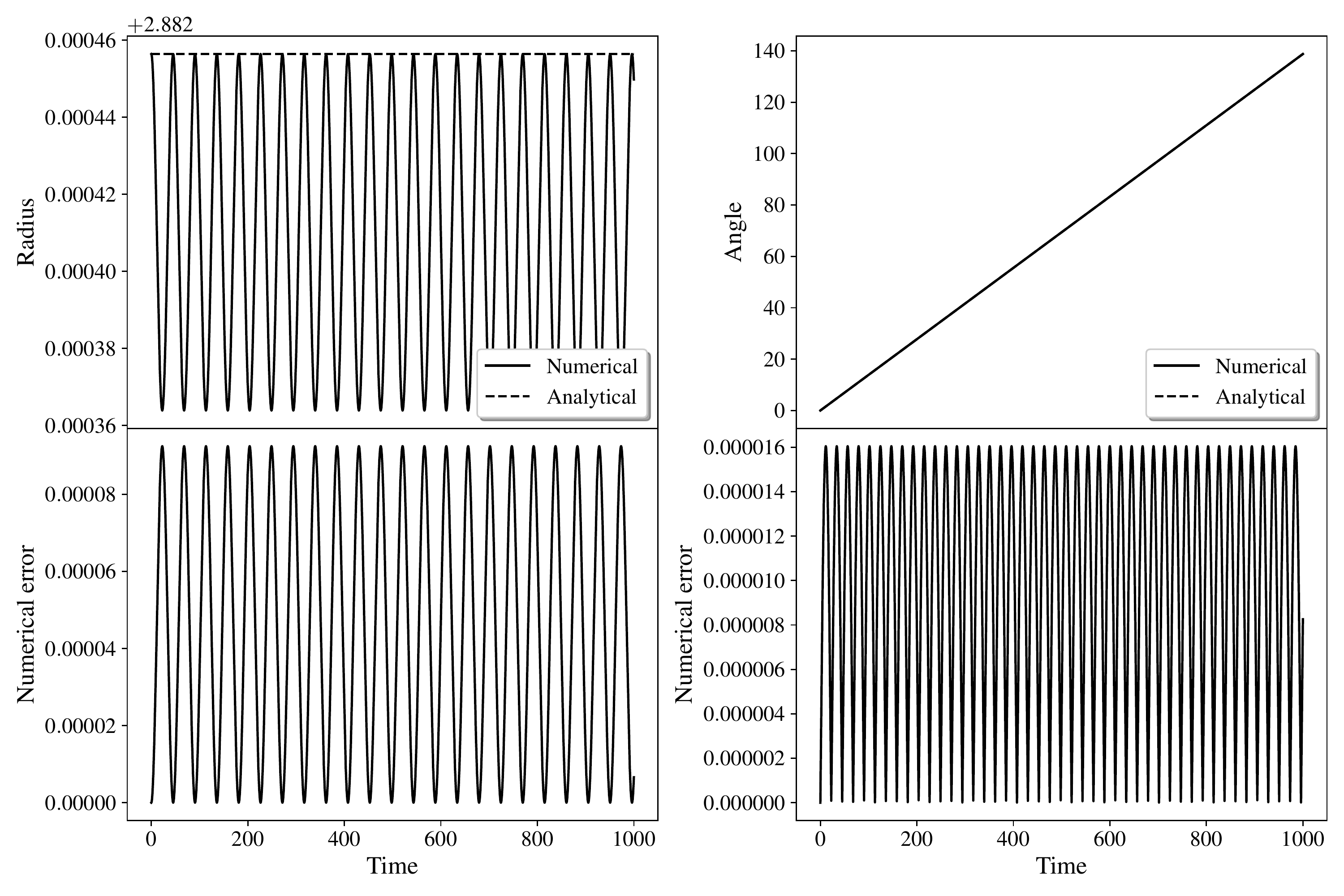} 
	\end{center}
	\caption{Relative error on energy as function of time. }
	\label{fig:consB_pos_verlet}
\end{figure*}

\begin{figure*}[t]
	\begin{center}
		\includegraphics[width=1.0\textwidth]{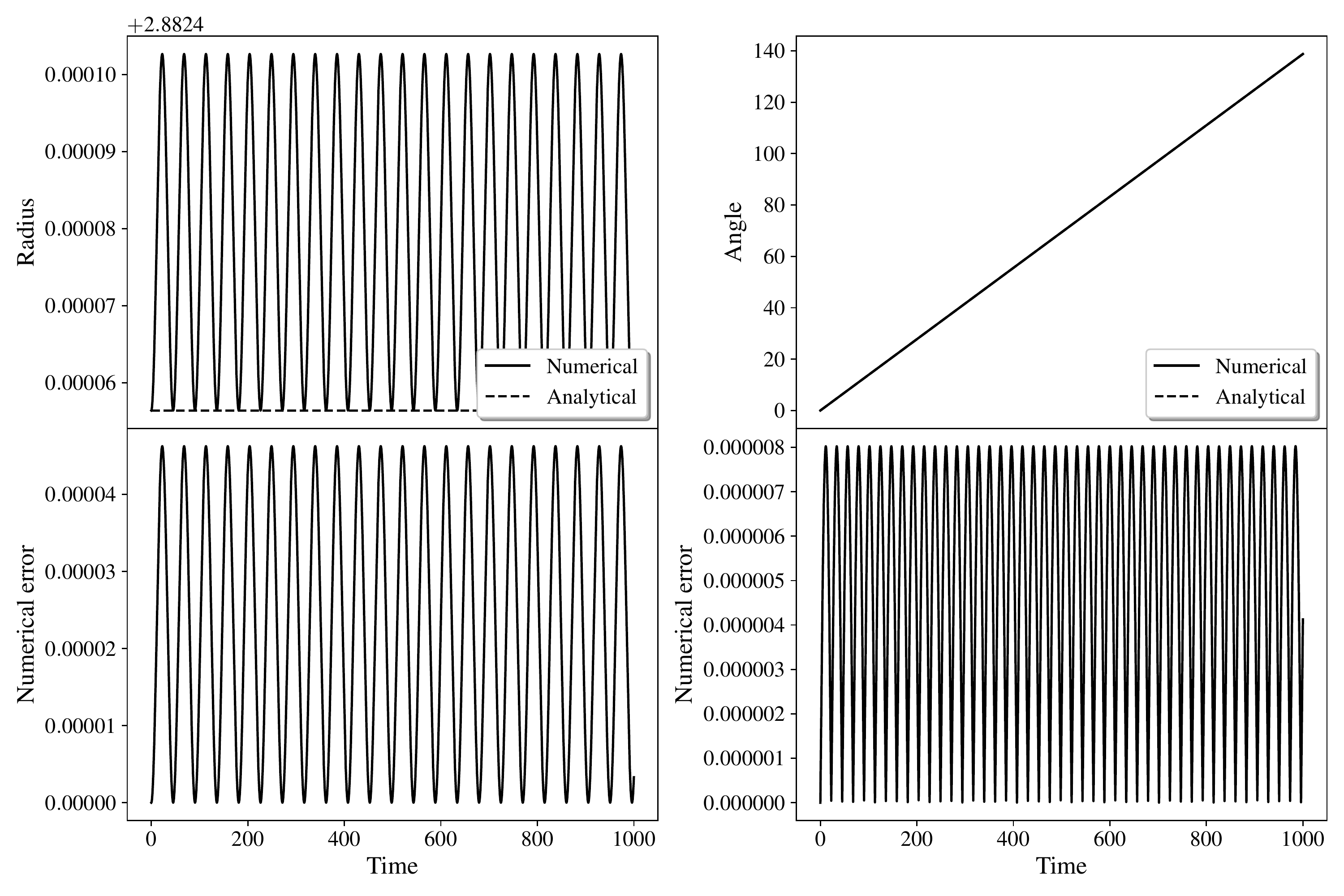} 
	\end{center}
	\caption{Relative error on energy as function of time. }
	\label{fig:consB_pos_leap}
\end{figure*}

In the second test, the actual error on position is evaluated. For these calculations, the final time is chosen as $t_{\mathrm{final}} = 1000$ and the number of time step is $N=10000$. The radius and angle are evaluated from the Cartesian components and the error on the radius and angle is simply defined as $\epsilon_{r} = |r(t) - r_{g}|$ and $\epsilon_{\theta} = |\theta(t) - \theta_{\mathrm{exact}}(t)|$. The results are displayed in Figs. \ref{fig:consB_pos_boris}, \ref{fig:consB_pos_verlet} and \ref{fig:consB_pos_leap} for the Boris, the Verlet-like and the Leapfrog-like schemes, respectively. The first observation is that all the numerical methods reproduce the analytical result with fairly high accuracy. However,  two conclusions can be reached by looking at the error in the numerical results. On the one hand, the accuracy of the Boris scheme for the radius of the trajectory is far superior than the other two schemes. Indeed, the error of the Boris scheme reaches machine precision ($\epsilon_{r} \approx 3.0 \times 10^{-14}$) while the error for the other two schemes oscillates, bounded by $\epsilon_{r} \lesssim 0.9 \times 10^{-4}$. On the other hand, the numerical error on the angle accumulates linearly in the Boris scheme, consistent with the findings of Ref. \cite{Ripperda_2018}, and can reach relatively high value in long term simulations. This linear accumulation of error is not observed for the Verlet and Leapfrog-like scheme. Rather, the error oscillates but stays bounded by $\epsilon_{\theta} \lesssim 0.16 \times 10^{-4}$. This interesting property is likely due to the volume-preserving properties of the numerical scheme. 

It was also observed (not shown here for simplicity) that the bound on the error can be lowered by increasing the number of time steps and decreasing $\Delta t$, as expected from the numerical method convergence rates. However, when one reaches a large number of time steps ($N \gtrsim 100000$), a small error starts accumulating, possibly due to the repeated third-order error at every iteration. This phenomenon has also been reported in simulations using other numerical methods \cite{petri_2017}. Nevertheless, the fact that the error stays bounded for both $r$ and $\theta$ when $\Delta t$ is not too small, makes the Verlet and Leapfrog schemes interesting alternatives for long term simulations.

\subsection{Plane wave}

The final test is for a particle immersed in a plane wave propagating in the $z$-direction. For an analysis of the spin dynamics of electrons in laser fields, see Ref. \cite{PhysRevA.87.052107}. This illustrates spin dynamics and the convergence of the numerical methods when the electromagnetic field is space-dependent. In particular, the electromagnetic field is given by
\begin{align}
\boldsymbol{E}(t,z) = \left(E\cos(\varphi),0,0\right) ,\\
\boldsymbol{B}(t,z) = \left(0,E\cos(\varphi),0\right) ,
\end{align}
where $E$ is the field amplitude and $\varphi = \omega(z-t)$, with $\omega$ the angular frequency. Remarkably, it is possible to find an exact solution of the Lorentz-BMT equation in such field configuration. The positions are given by \cite{saxena1993}
\begin{align}
x(t) &= \frac{qE}{m\omega^{2}} \left[1 - \cos (\varphi)\right], \\
y(t) &= 0, \\
z(t) &= \frac{q^{2}E^{2}}{8m^{2}\omega^{3}} \left[\sin(2\varphi) - 2\varphi\right].
\end{align}
On the other hand, the spin dynamics can be extracted from the matrix spinor via \eqref{SpinDef}.

\begin{figure*}[t]
	\begin{center}
		\includegraphics[width=1.0\textwidth]{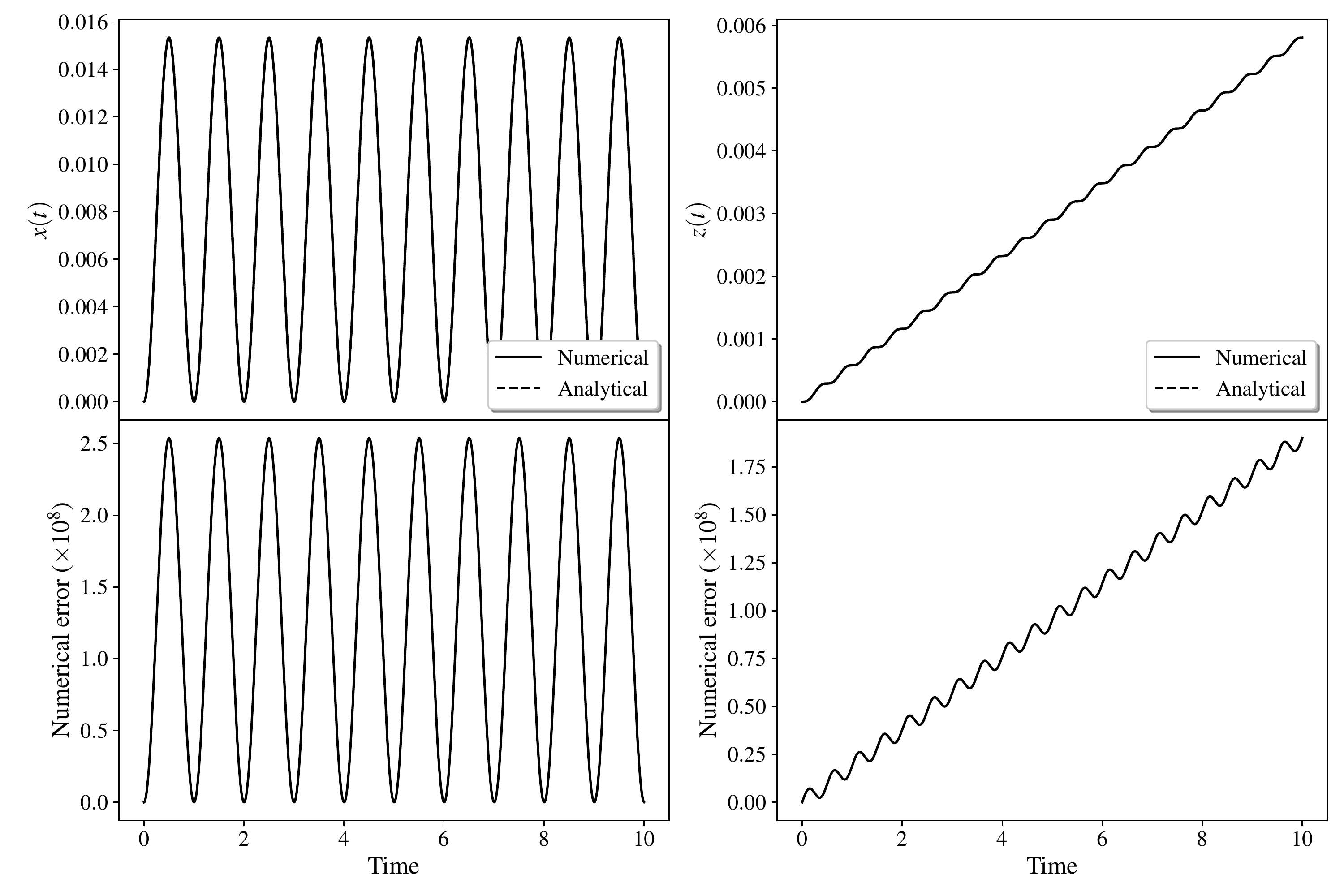} 
	\end{center}
	\caption{Relative error on energy as function of time. }
	\label{fig:plane_pos_boris}
\end{figure*}

\begin{figure*}[t]
	\begin{center}
		\includegraphics[width=1.0\textwidth]{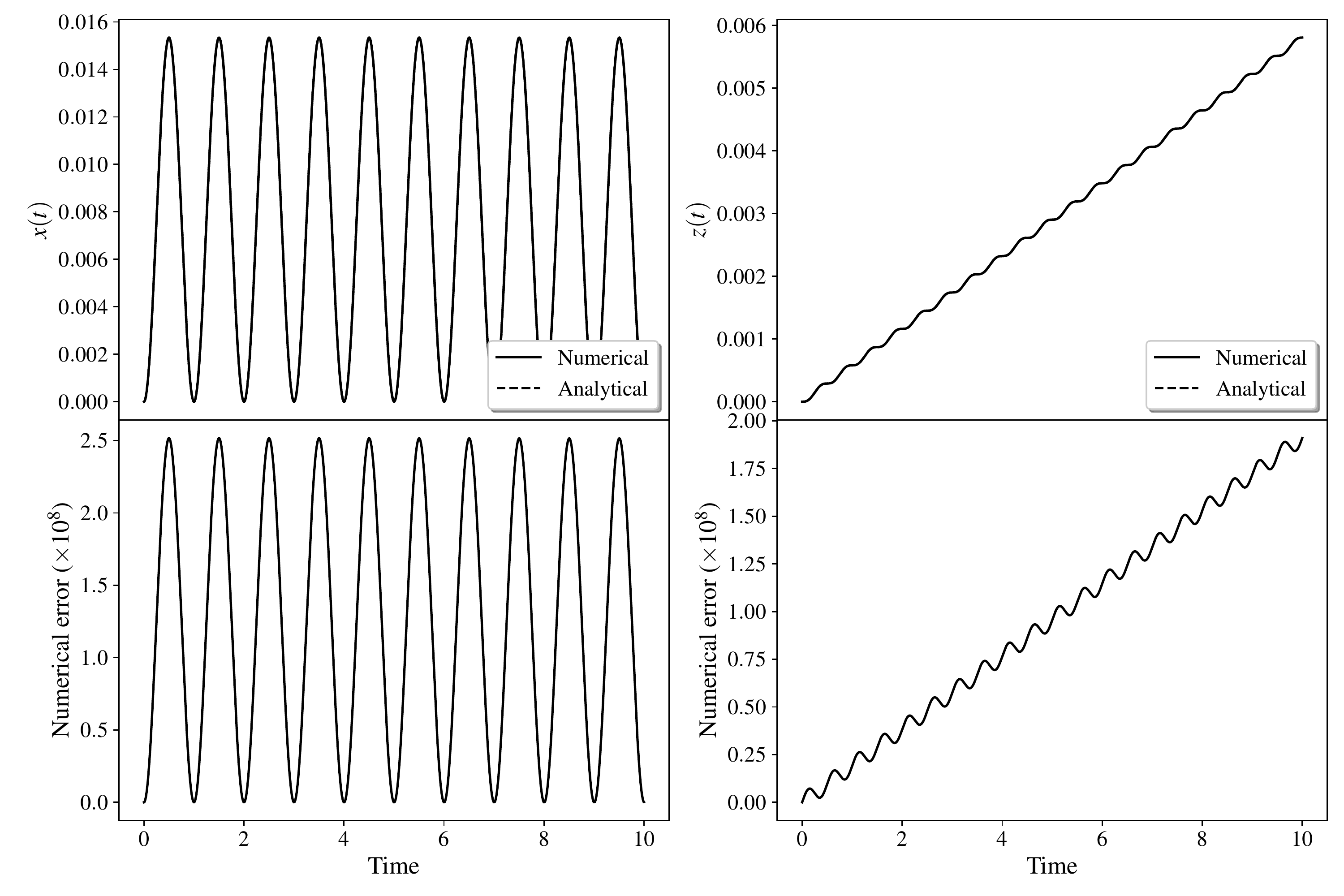} 
	\end{center}
	\caption{Relative error on energy as function of time. }
	\label{fig:plane_pos_verlet}
\end{figure*}

In the simulations for a plane wave electromagnetic field, the final time is $t_{\mathrm{final}} = 10$ while the number of time steps is $N=10000$, making for $\Delta t = 0.001$. The electric field strength is set to $E=1$ while the angular frequency is  $\omega = 2\pi$. The particle is initially at the origin $\boldsymbol{x}(0) = 0$ and at rest $\boldsymbol{v}(0) = 0$. The comparison with the analytical solution is displayed in Figs. \ref{fig:plane_pos_boris} and \ref{fig:plane_pos_verlet}, for the Boris and Verlet-like methods, respectively. The results for the Leapfrog-like scheme are not shown for simplicity and because they are similar to the Verlet scheme. The numerical error is evaluated using Eq. \eqref{eq:error} and presented below the position. These numerical results demonstrate that the Boris and Verlet-like methods reproduce the exact solution accurately and perform equally well, both having numerical errors bounded by $\epsilon \lesssim 2.5 \times 10^{-8}$ for the $x$ and $z$ positions. 

\begin{figure*}[t]
	\begin{center}
		\includegraphics[width=1.0\textwidth]{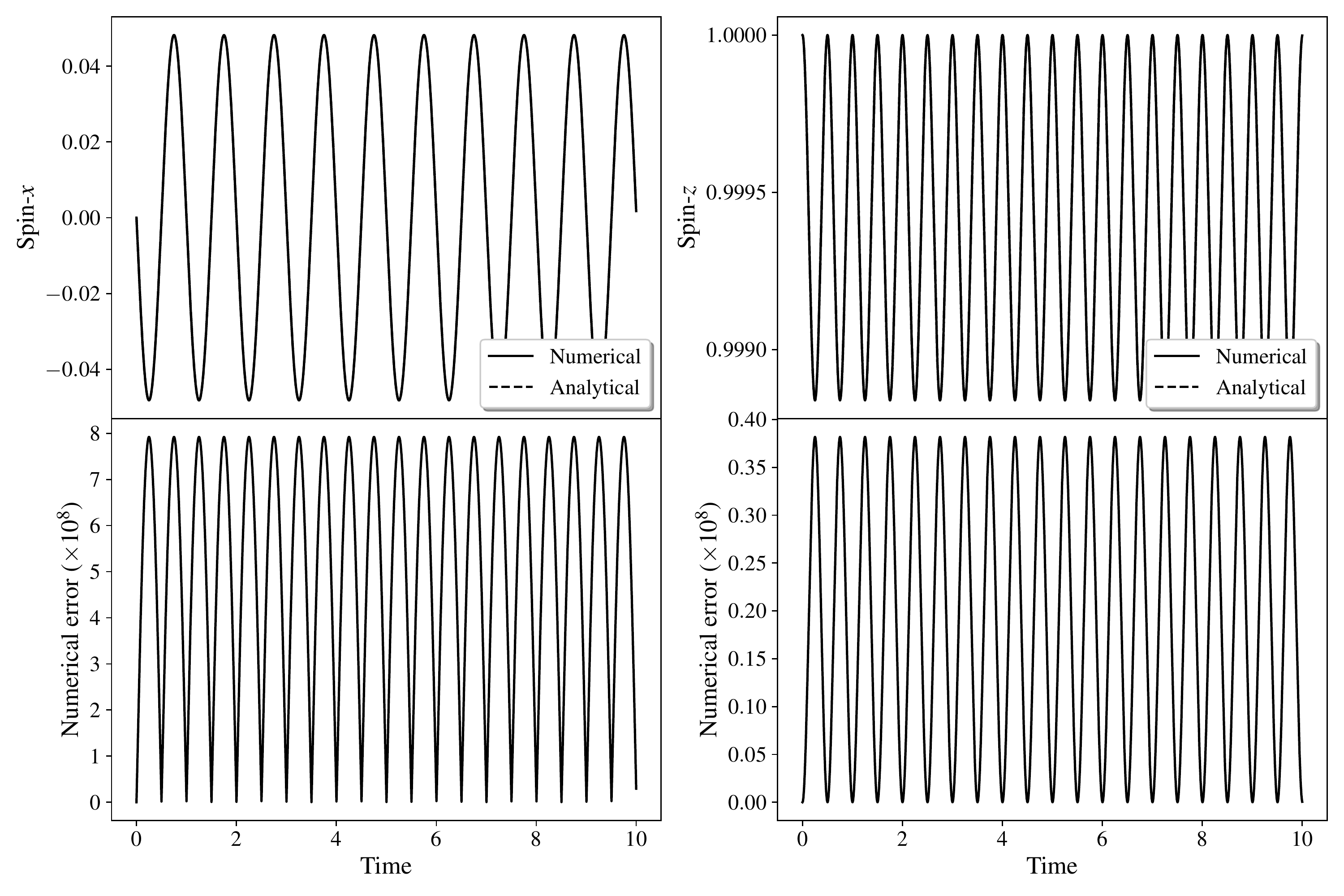} 
	\end{center}
	\caption{Relative error on energy as function of time. }
	\label{fig:plane_spin_verlet}
\end{figure*}

However, the main advantage of the Verlet-like scheme is that spin dynamics can be obtained easily via the relation \eqref{SpinDef}. The numerical results are compared to this analytical solution is Fig. \ref{fig:plane_spin_verlet} for the Verlet-like scheme (again, the Leapfrog method is not displayed because it presents similar results). The numerical error is also evaluated for each spin component using $\epsilon_{\mathrm{s}} = |S_{i}^{N} - S_{i,\mathrm{exact}}|$, for $i=x,z$. Again, the numerical method reproduce the exact solution very accurately, with numerical errors bounded by $\epsilon_{\mathrm{s}} \lesssim 8.0 \times 10^{-8}$ for both spin components. 

\section{Conclusion}\label{Conclusion}

In this work, the Lorentz-BMT system of equations was solved numerically in the Clifford algebra representation. Two numerical schemes were developed and tested against the Boris pusher by comparing with analytical solutions. It was demonstrated that the numerical schemes have bounded numerical errors, even in strong magnetic fields, in contrast to the Boris method. This long term accuracy is attributed to their volume-preserving properties. We also showed that the Verlet- and Leapfrog-like schemes share the strengths of the Boris pusher: they have second order convergence, they are explicit, they preserve energy when there is no electric field and finally, they are simple and easy to implement. In addition, they can be used to obtain the spin dynamics of the charged particle, without solving explicitly another differential equation. All of these features make them very appealing for applications in plasma physics, accelerator physics, astrophysics and others. 

The numerical methods developed in this article could be improved in several ways. For instance, the split operator method can be extended to third order accuracy and even higher \cite{Bandrauk_1994,130003901667,Wiebe_2010}. Combining these results on exponential operators with usual methods for solving ordinary differential equations, we conjecture that higher order numerical schemes could be obtained. This will be the topic of future work.

\section*{Acknowledgments}
RC and AGC contributed equally to this work. D.I.B. is supported by Air Force Office of Scientific Research (AFOSR) (grant FA9550-16-1-0254), Army Research Office (ARO) (grant W911NF-19-1-0377). The views and conclusions contained in this document are those of the authors and should not be interpreted as representing the official policies, either expressed or implied, of AFOSR, ARO, or the U.S. Government. The U.S. Government is authorized to reproduce and distribute reprints for Government purposes notwithstanding any copyright notation herein. This research was enabled in part by support provided by Calcul Qu\'{e}bec (www.calculquebec.ca) and Compute Canada (www.computecanada.ca).

\appendix

\section{Volume-preservation of the dynamic equations \label{app:vol_ode}}

In this appendix, it is demonstrated that the ODE system given in Eqs. \eqref{eq:Lambda_lab} and \eqref{eq:x_lab} is volume preserving. For this purpose, we follow Ref. \cite{quispel2001six} and introduce similar notation. First, the ODE system is written in the general form
\begin{align}
\label{eq:ode_sys}
    \frac{dz(t)}{dt} = G(z), \;\; \mbox{for} \;\; z,G \in \mathbb{C}^{m},
\end{align}
where $z = (z_{1}, \cdots, z_{m})^{T}$ is a real vector containing the ODE degrees of freedom (DOF) while $G$, also a vector, specifies the dynamics of all DOF. The exact flow $\varphi_{\Delta t}$ of the ODE system is defined by
\begin{align}
z(t + \Delta t) = \varphi_{\Delta t}(z(t)).
\end{align}
To write the ODE system \eqref{eq:Lambda_lab}-\eqref{eq:x_lab} in the form of Eq. \eqref{eq:ode_sys}, the vectorization operator is introduced
\begin{align}
\vec{A}:= \mathrm{vec}(A) = \nonumber \\
[A_{11}, \cdots,  A_{m1}, A_{12}, \cdots A_{m2}, \cdots, A_{1m}, \cdots, A_{mm}]^{T},
\end{align}
for any $m$-by-$m$ matrices $A$ with components $(A_{ij})_{i,j=1,\cdots,m}$. This operation is a mapping $\mathrm{vec}: \mathrm{M}_{m}(\mathbb{C}) \rightarrow  \mathbb{C}^{m^{2}}$ that transforms a square matrix into a vector. This operation obeys some properties, in particular $\mathrm{vec}(AB) = (\mathbb{I}_{m} \otimes A) \mathrm{vec}(B)$. Armed with this notation, it is now possible to demonstrate that the ODE system  is volume-preserving. 

An ODE system is divergence-free when
\begin{align}
\label{eq:div}
\sum_{i=1}^{m} \frac{\partial G_{i}}{\partial z_{i}} = 0.
\end{align}
In addition, a divergence-free ODE system is volume preserving \cite{quispel2001six}, thus we now demonstrate that Eq. \eqref{eq:div} holds for \eqref{eq:Lambda_lab} and \eqref{eq:x_lab}. First, the vectorization mapping is applied to the ODE system yielding
\begin{align}
\frac{d\vec{\Lambda}}{dt}  &= \frac{q}{2m \gamma } (\mathbb{I}_{2} \otimes F) \vec{\Lambda} := G^{(\Lambda)}, \\
\frac{d\vec{x}}{dt} &= \frac{\mathrm{vec}\left( \Lambda \Lambda^{\dagger} \right)}{\gamma} := G^{(x)}.
\end{align}
where $\vec{\Lambda},\vec{x}, G^{(x)}, G^{(\Lambda)} \in \mathbb{C}^{4}$ are four-dimensional vectors, obtained from the stacking of matrix components.   
Obviously, the derivatives 
\begin{align}
\frac{\partial G^{(x)}{i}}{\partial \vec{x}_{i}} = 0  \;\; \mbox{for} \;\; i=1, \cdots, 4,
\end{align}
simply because $G^{(x)}$ has no explicit dependence on $x$. 
On the other hand, derivatives of $G^{(\Lambda)}$  are not zero, rather we have
\begin{align}
\frac{\partial G^{(\Lambda)}_{1}}{\partial \vec{\Lambda}_{1}} &= \frac{\partial G^{(\Lambda)}_{3}}{\partial \vec{\Lambda}_{3}} = E^{3} + i B^{3}, \\ 
\frac{\partial G^{(\Lambda)}_{2}}{\partial \vec{\Lambda}_{2}} &= \frac{\partial G^{(\Lambda)}_{4}}{\partial \vec{\Lambda}_{4}} = -E^{3} - i B^{3}.
\end{align}
However, when taking the divergence, a sum on all these contributions is taken and we get $\nabla \cdot G^{(\Lambda)} = 0$. As a consequence, the ODE system is divergence-free, implying that it is also volume-preserving \cite{quispel2001six}. It should be noted here that the phase space volume considered here is for the ODE system, it is different from the one for a Hamitonian system $(\boldsymbol{x},\boldsymbol{p})$. However, it is clearly an intrinsic property of the ODE system that needs to be fulfilled by numerical schemes.

\section{Volume-preservation of the numerical schemes \label{app:vol_num_schemes}}

We now turn to proving that the Verlet-like scheme is manifestly volume preserving. First, the vectorization operation is applied and the scheme is written as
\begin{align}
\label{eq:phi1}
\phi^{n+\frac{1}{2}} &:=
\begin{cases}
\vec{x}^{n+\frac{1}{2}} = \vec{x}^{n} + \frac{\Delta t}{2 \gamma^{n}} \mathrm{vec}(\Lambda^{n}\Lambda^{n \dagger}) \\
\vec{\Lambda}^{n+\frac{1}{2}} = \vec{\Lambda}^{n}
\end{cases} , \\
\label{eq:phi2}
\phi^{\tilde{n}} &:=
\begin{cases}
\vec{x}^{\tilde{n}} = \vec{x}^{n+\frac{1}{2}} \\
\vec{\Lambda}^{\tilde{n}} = (\mathbb{I}_{2} \otimes U^{n}) \vec{\Lambda}^{n+\frac{1}{2}}
\end{cases} , \\
\label{eq:phi3}
\phi^{n+1} &:=
\begin{cases}
\vec{x}^{n+1} = \vec{x}^{\tilde{n}} + \frac{\Delta t}{2 \gamma^{\tilde{n}}} \mathrm{vec}(\Lambda^{\tilde{n}}\Lambda^{\tilde{n} \dagger}) \\
\vec{\Lambda}^{n+1} = \vec{\Lambda}^{\tilde{n}} 
\end{cases},
\end{align}
where $\phi^{n}$ are approximated flows and $\tilde{n}$ denotes an intermediary time step.
Setting 
\begin{align}
z = 
\begin{bmatrix}
\vec{x} \\
\vec{\Lambda}
\end{bmatrix},
\end{align}
the Jacobian of the flow can be written in matrix form as
\begin{align}
\frac{\partial \phi^{n+\frac{1}{2}}}{\partial z^{n}} &= 
\begin{bmatrix}
\mathbb{I}_{4} & M^{(1)} \\
0 & \mathbb{I}_{4}
\end{bmatrix} ,\\
\frac{\partial \phi^{\tilde{n}}}{\partial z^{n+\frac{1}{2}}} &= 
\begin{bmatrix}
\mathbb{I}_{4} & 0 \\
M^{(2)} & \mathbb{I}_{2} \otimes U^{n}
\end{bmatrix}, \\
\frac{\partial \phi^{n+1}}{\partial z^{\tilde{n}}} &= 
\begin{bmatrix}
\mathbb{I}_{4} & M^{(3)} \\
0 & \mathbb{I}_{4}
\end{bmatrix} ,
\end{align}
where the four-by-four matrices $M^{(1,3)}$ comes from the derivative with respect to $\Lambda$ in Eqs. \ref{eq:phi1} and \ref{eq:phi3}, while the matrix $M^{(2)}$ comes from the derivative with respect to $x$ in Eq. \ref{eq:phi2}. Their explicit expression is not important because we are interested in the determinant of the Jacobian. Using the properties of determinant, the latter are given by
\begin{align}
\mathrm{det}\left( \frac{\partial \phi^{n+\frac{1}{2}}}{\partial z^{n}} \right) &= 1 \\
\mathrm{det}\left(\frac{\partial \phi^{\tilde{n}}}{\partial z^{n+\frac{1}{2}}} \right) &= 1 \\
\mathrm{det}\left(\frac{\partial \phi^{n+1}}{\partial z^{\tilde{n}}} \right) &= \mathrm{det}\left(U^{n}\right)^{2}.
\end{align}
The last determinant can be evaluated from the definition of $U^{n}$ and the identity for the determinant of a matrix exponential:
\begin{align}
\mathrm{det}\left(U^{n}\right) = \exp \left[ \Delta t \frac{q}{2m \gamma^{n+\frac{1}{2}}} \mathrm{Tr}(F(x^{n+\frac{1}{2}}))  \right] .
\end{align}
However, from the definition of the electromagnetic field, we have that $\mathrm{Tr}(F(x^{n+\frac{1}{2}})) = 0$, confirming that the last determinant is also unity. This concludes the demonstration that the Verlet-like scheme is volume preserving.

The argument for the leapfrog scheme is very similar. The approximated flow is now
\begin{align}
\label{eq:phi1_leap}
\phi^{n+\frac{1}{2}} &:=
\begin{cases}
\vec{x}^{n+\frac{1}{2}} = \vec{x}^{n} + \frac{\Delta t}{ \gamma^{n}} \mathrm{vec}(\Lambda^{n}\Lambda^{n \dagger}) \\
\vec{\Lambda}^{n+\frac{1}{2}} = \vec{\Lambda}^{n}
\end{cases} , \\
\label{eq:phi2_leap}
\phi^{n+1} &:=
\begin{cases}
\vec{x}^{n+1} = \vec{x}^{n+\frac{1}{2}} \\
\vec{\Lambda}^{n+1} = (\mathbb{I}_{2} \otimes U^{n}) \vec{\Lambda}^{n+\frac{1}{2}}
\end{cases}  ,
\end{align}
with the understanding that $x$ and $\Lambda$ are staggered. In matrix form, the Jacobian of the flow gives
\begin{align}
\frac{\partial \phi^{n+\frac{1}{2}}}{\partial z^{n}} &= 
\begin{bmatrix}
\mathbb{I}_{4} & \tilde{M}^{(1)} \\
0 & \mathbb{I}_{4}
\end{bmatrix} ,\\
\frac{\partial \phi^{n+1}}{\partial z^{n+\frac{1}{2}}} &= 
\begin{bmatrix}
\mathbb{I}_{4} & 0 \\
M^{(2)} & \mathbb{I}_{2} \otimes U^{n}
\end{bmatrix}.
\end{align}
The last steps of the proof are the same as for the Verlet-like scheme and we obtain a unit determinant. Thus, we conclude that the leapfrog scheme is also volume preserving. 

\bibliography{bibliography}

\end{document}